\documentclass[12pt]{article}
\pdfoutput=1
\usepackage{setspace}
\usepackage[top=1in, bottom=1in, left=1in, right=1in]{geometry}
\usepackage{enumitem} 
\setlist{nolistsep} 

\usepackage[colorlinks,linkcolor=blue,citecolor=blue,urlcolor=blue]{hyperref}
\usepackage{mathtools,amssymb,amsthm,amsthm,latexsym,bbm}
\usepackage{amscd,amsxtra} 
\usepackage{graphics,graphicx,xcolor}
\usepackage{ifpdf}
\usepackage[caption=false]{subfig}
\usepackage{multirow}

\usepackage[authoryear]{natbib}
\usepackage{comment}
\usepackage{ulem}
\usepackage{authblk} 

\theoremstyle{definition}

\newtheorem*{defn*}{Definition}

\newtheorem{assume}{Assumption}
\theoremstyle{remark}

\newtheorem{example}{Example}

\def\xv{\boldsymbol x}

\def\Xv{\boldsymbol X}

\newcommand{\Ac}{\mathcal{A}}

\newcommand{\Pc}{\mathcal{P}}

\newcommand{\Sc}{\mathcal{S}}

\newcommand{\Wc}{\mathcal{W}}

\newcommand{\alphav}{\mbox{\boldmath{$\alpha$}}}

\newcommand{\deltav}{\mbox{\boldmath{$\delta$}}}

\newcommand{\muv}{\mbox{\boldmath{$\mu$}}}

\def\1v{\mathbf 1}
\def\0v{\mathbf 0}
\def\Id{\mathbf I} 

\newcommand{\Ind}[1]{\mathbbm{1}_{\left\{ {#1} \right\} }}

\newcommand{\R}{\mathbb R}

\newcommand{\ds}{\displaystyle}
\newcommand{\mb}{\mbox}
\newcommand{\wh}{\widehat}
\newcommand{\argmin}{\operatornamewithlimits{argmin}}

\newcommand{\set}[1]{\left\{#1\right\}}

\def\ie{\textit{i.e.}}
\def\eg{\textit{e.g.}}

\title{Significance Analysis for Pairwise Variable Selection in Classification}

\author{Xingye Qiao\thanks{Corresponding author}}
\affil{Department of Mathematical Sciences\authorcr  State University of New York, Binghamton, NY 13902-6000.\authorcr E-mail: \texttt{qiao@math.binghamton.edu}}

\author{Yufeng Liu and J. S. Marron}
\affil{Department of Statistics and Operations Research\authorcr
University of North Carolina at Chapel Hill, Chapel Hill, NC 27599.\authorcr
E-mail: \texttt{(yfliu,marron)@email.unc.edu}

}
\date{}
\begin{document}
\maketitle

\newpage

\begin{abstract}
The goal of this article is to select important variables that can distinguish one class of data from another.  A marginal variable selection method ranks the marginal effects for classification of individual variables, and is a useful and efficient approach for variable selection. Our focus here is to consider the bivariate effect, in addition to the marginal effect. In particular, we are interested in those pairs of variables that can lead to accurate classification predictions when they are viewed jointly. To accomplish this, we propose a permutation test called Significance test of Joint Effect (SigJEff). In the absence of joint effect in the data, SigJEff is similar or equivalent to many marginal methods. However, when joint effects exist, our method can significantly boost the performance of variable selection. Such joint effects can help to  provide additional, and sometimes dominating, advantage for classification. We illustrate and validate our approach using both simulated example and a real glioblastoma multiforme data set, which provide promising results.
\end{abstract}

\newpage
\section{Introduction}\label{sec:intro}
In many real data applications, such as bioinformatics and medical image analysis, there are
thousands to hundreds of thousands variables available for modeling (\ie, $\Xv \in \R^d$, where $d\approx 10^3$ to $10^5$). It is often, however, that only a small number of them truly influence the response variable $Y$. The aim of variable selection is to
identify these variables which strongly influence the response variable and thus have great
predictive power. Variable selection plays a fundamental role in high-dimensional statistical
inference. In this article, our focus is on variable selection for the binary classification problem where the response variable $Y\in\{+1,-1\}$.

Our emphasis in this paper is on genetic applications, where each gene is a variable. However, the lessons are
broadly applicable. Classification based on gene expression data has been shown useful in cancer research \citep{Golub1999Molecular}. \textit{Marginal}, \ie, gene-by-gene, methods assess each individual gene  separately. Such methods help to identify genes that are important marginally, and eliminate genes that are almost \textit{useless} in the marginal sense. There is a large literature in this direction. An incomplete list includes methods based on the classical two-sample $t$-statistic which can be seen in most statistical textbooks, \eg, \citet{Peck2011Statistics:}; the Empirical Bayes approach \citep{Efron2001Empirical}; the Significance Analysis of Microarray \citep[SAM;][]{Tusher2001Significance}; a mixture model approach \citep{Pan2003mixture}, among others. There are other methods based on some improved marginal statistics, such as \citet{Baldi2001Bayesian}, \citet{Zuber2009Gene}, \citet{Wu2005Differentiala,Wu2005Differential}, \textit{etc}. \citet{Pan2002comparative} has compared several of these methods and concluded that the statistics involved are similar, although they may adopt different assumptions. Recently, \citet{Fan2008Sure} provides some theoretical justification for marginal screening in a regression setting. In this article, we take SAM \citep{Tusher2001Significance} as a representative example of the marginal methods. In SAM, a significance test is implemented for each variable individually and a list of statistics (denoted as $t_i$) for these tests is obtained. A threshold is then calculated for an overall error control so that all the variables with $|t_i|$ greater than the threshold are claimed as \textit{being significant}. Marginal methods are often efficient in computation, and useful in various real application cases. However, important joint effects among variables may be missed by these marginal procedures.

\begin{figure*}[!ht]
		\includegraphics[width=\linewidth]{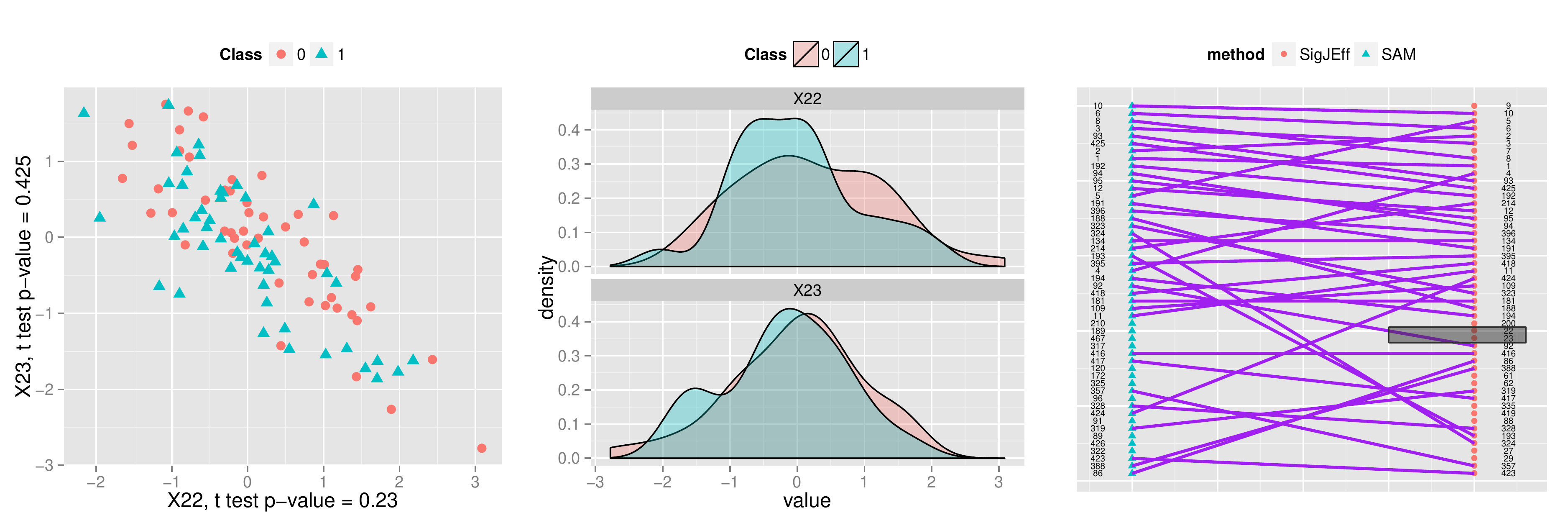}
		\caption{Left panel: 2D scatter plot of the data from the AR(1) simulation example (see the setting in Section \ref{sec:settings}) based on variable \#22 and variable \#23. Two-sample $t$-tests are conducted for both variables and both tests yield insignificant $p$-values, annotated in the axis labels. See also the middle panel, where the density estimations for both classes are plotted for each variable (the top subplot and the bottom subplot). The two classes demonstrate no significant difference on either variable. However, when these two variables are viewed together, a pattern can be seen that the Class `1' samples are around the southwest corner of the scatter plot while the Class `0' samples are at the northeast. Such a pattern is not visible when we consider each individual variable only. Right panel: The variable indices of the first 50 variables that are selected by SigJEff (on the right) and by SAM (on the left) respectively. This shows the different ranks of variables by the two methods. A variable that is selected by both methods is annotated by a purple line segment connecting its positions in both lists. The pair of variables \#22 and \#23 are not selected by SAM, but selected by SigJEff due to their joint effect.}
		\label{fig:sim1_plot_intro}
\end{figure*}

In practice, it can be reasonable to assume that many genes interact with each other, and they work together to drive the variation in a particular phenotype. For the general application of classification, it is possible to have two variables with weak marginal effects that work together to yield a strong joint effect for classification. In this case, marginal screening methods can be insufficient since these Alone-Nonsignificant-Together-Significant (ANTS) variables may not be selected due to their weak marginal signals, even though the joint effects can be substantial. For this reason, an approach which incorporates the joint effect information can potentially select more useful variables.

In practice, due to the constrained research budget, researchers may not be able to look into all the potentially important variables. The common practice is to provide a sorted list of variables, from the most important to the least, for researchers to choose according to their budget. Marginal methods rank variables based on their marginal effects. Thus, those ANTS variables may receive a lower priority and is more likely to be missed. We propose to rank the variables based on a new criterion on the joint effect, instead of the marginal effect of a variable. Because of the use of this new criterion, ANTS variables in general would receive a better rank than that by a marginal method. The right panel in Figure \ref{fig:sim1_plot_intro} shows the rankings of variables from a simulated dataset (details in Section \ref{sec:settings}), given by SAM and our proposed procedure, \textit{Significance test of Joint Effects} (SigJEff). We can clearly see the common places and differences between these two rankings: some most important variables are selected by both methods with the highest priorities, while some less SAM-important variables can receive better ranks when SigJEff is used. In the left panel, we examine some typical variables, variables \#22 and \#23 (contained in the shadow box in the right panel), by drawing the scatter plot of the data set based on them. From the left panel, we can see that the two variables are truly ANTS variables, namely, alone nonsignificant but together significant. In particular, two-sample $t$-tests are conducted for both variables and both tests yield insignificant $p$-values. This can be also seen from the middle panel, where the density estimations for both classes are plotted for each variable. The two classes demonstrate no significant difference on either variable. However, when these two variables are viewed together, a pattern can be seen that the Class `1' samples are around the southwest corner of the scatter plot while the Class `0' samples are at the northeast. Such a pattern is not visible when we consider each individual variable only. More details of this example will be analyzed in Section \ref{sec:settings}.

In this article, we conduct the significance analysis for  pairwise variable selection in classification. Here the term \textit{pairwise} refers to pairs of variables, instead of pairs of observations or pairs of classes.  Our proposed method SigJEff is based on a permutation approach that assesses the joint effect of a variable pair. Specifically, we want to assess whether there is statistically significant difference between two classes based on these two variables only.

The rest of the article is organized as follows. In Section \ref{sec:method}, we introduce the
SigJEff procedure. Section \ref{sec:computing} is devoted to the computation of our method. Section \ref{sec:simulation} presents the numerical properties of our method, including three different simulation settings. A real data application study is conducted in Section \ref{sec:realdata}. In addition to the marginal method SAM, we also compare our approaches with those from \textit{Least Absolute Shrinkage and Selection Operator}
\citep[LASSO;][]{Tibshirani1996Regression}. Concluding remarks are provided in Section \ref{sec:Conclusion}.

\section{Methodology}\label{sec:method}
The proposed SigJEff procedure is a permutation procedure to assess joint effects between pairs of variables. In high dimensional problems, permutating all pairs of variables would be computational costly. Moreover, the resulting statistics of the pairs can be highly correlated, which makes it difficult to control false discoveries. To overcome these difficulties, in this paper, we first partition the $d$ variables to $\left\lfloor d/2\right\rfloor$ disjoint pairs, where $\left\lfloor t\right\rfloor$ is the largest integer less than or equal to $t$. Once the partition is done, we will conduct a permutation test for each given pair in the partition. Lastly, a $p$-value is calculated for each pair, and a sorted list of pairs of variables will be given.

\subsection{Variable partition}\label{sec:partition}
Let $\deltav = \wh{\muv_1} - \wh{\muv_2}$ be the sample mean difference between the two classes and $\wh{\Sigma}$ the within-class sample covariance matrix. In general, for a variable set $S\subset\set{1,\cdots,d}$, a vector $\alphav\in\R^d$ and a matrix $A\in\R^{d\times d}$, let $\alphav_S$ and $A_{S,S}$ denote the subvector of $\alphav$ and the principal submatrix of $A$ corresponding to $S$ respectively.

Let $M_{d\times d}=\left(m_{(i,j)}\right)$ be a symmetric matrix whose off-diagonal $(i,j)$th entry is the Mahalanobis distance between the two classes based on variables $i$ and $j$, \ie
\begin{align*}
	m_{i,j} = \begin{cases}
		\deltav_{(i,j)}'\left(\wh{\Sigma}_{(i,j),(i,j)}\right)^{-1}\deltav_{(i,j)},~&i<j,\\
		0,~&i=j,\\
		m_{j,i},~&i>j.
	\end{cases}
\end{align*}

Without loss of generality, we assume that the dimension $d$ is an even number. To this end we conduct a partition of the $d$ variables into $(d/2)$ pairs as follows. We first capture the pair of variables $(i,j)$ with the greatest $m_{i,j}$. Then, we delete the $i$th and $j$th rows and $i$th and $j$th columns from the $M$ matrix. Among all the remaining $d-2$ variables, we then find the pair $(i,j)$ with the greatest $m_{i,j}$ in the remaining $M$ matrix. This procedure continues until all the pairs, thus all the variables, are captured. Note that to carry out this partition, $d(d-1)/2$  Mahalanobis distances need to be calculated. However, this is the only time in the whole procedure for us to calculate all the distances for all $d(d-1)/2$ pairs of variables. The practice of partitioning the variables serves to reduce the computational cost for the permutation stage of our analysis later.

\subsection{Pair selection by permutation}
Once we have identified the $\left\lfloor d/2\right\rfloor$ pairs of variables, for each pair, we conduct a permutation test by randomly relabeling the class labels $Y$. Let $\Pc = \set{(i,j)}$ be the partition we obtain in Section \ref{sec:partition}. For the $p$th permutation, for each $(i,j)\in\Pc$, the Mahalanobis distance $m_{i,j}^p$ is calculated for the permuted data. Note that here we do not conduct variable partition for the permuted data, nor do we calculate the Mahalanobis distances for all pairs. Instead, only the Mahalanobis distances for the pairs in the give partition $\Pc$ are calculated. In practice we choose the number of permutations $P=1000$.

A $p$-value will then be calculated. We allow three versions of the $p$-value in our implementation.
\begin{itemize}
	\item The empirical $p$-value is defined as $$\ds{p_{(i,j)} =\frac{1}{P} \sum_{p=1}^P \Ind{m_{i,j}^p>m_{i,j}}}.$$ This is the simplest version and is often used in practice. One potential drawback is that one may have ties due to the discrete nature of this $p$-value.
	\item The Gaussian fit $p$-value is calculated as $$p_{(i,j)} = 1-\Phi^{-1}\left[\left\{m_{i,j}-ave(m_{i,j}^p)\right\}/std(m_{i,j}^p)\right],$$where $\Phi^{-1}$ is the inverse of the cumulative distribution function of the standard normal distribution, $ave(\cdot)$ is the sample average and $std(\cdot)$ is the sample standard deviation.
	\item The robust Gaussian fit $p$-value is calculated as $$p_{(i,j)} = 1-\Phi^{-1}\left[\left\{m_{i,j}-median(m_{i,j}^p)\right\}/mad(m_{i,j}^p)\right],$$where $median(\cdot)$ is the sample median and $mad(\cdot)$ is the median absolute deviations, which are the robust counterparts of sample average and sample standard deviation.
\end{itemize}

If the number of permutations $P$ is large and if the effect is not very strong, these three versions of $p$-value will produce similar rankings for the pairs. However, when the joint effect in the data for certain pairs is very strong while $P$ is not large enough, then it is possible that the statistic calculated from the original data is greater than all its $P$ counterparts from the permuted data, in which case the empirical $p$-value is calculated as zero. In such cases, we suggest the use of either Gaussian fit version to obtain some approximated and interpretable assessment of the true $p$-value.

Lastly, we sort the $p$-values in an ascending order, set a threshold for the $p$-values, and claim significance for all the pairs of variables  with $p$-values less than or equal to the threshold.

\subsection{Algorithm}
We summarize the algorithm as follows.

\noindent\textbf{SigJEff Algorithm:}
\begin{enumerate}
				\item Partition:
				\begin{enumerate}
					\item Calculate $m_{i,j} = \deltav_{(i,j)}'\left(\wh{\Sigma}_{(i,j),(i,j)}\right)^{-1}\deltav_{(i,j)}$ for all $1\leq i<j\leq d$.
					\item Let $\Pc = \emptyset$.
					\item For $s$ = 1 to $\left\lfloor d/2\right\rfloor$,
							\begin{enumerate}
								\item Let $\Pc \leftarrow \Pc\cup \set{(i^*,j^*)}$, where $(i^*,j^*)=\ds{\argmin_{(i,j)}m_{i,j}}$.
								\item Let $m_{i^*,j'}\leftarrow0$, $m_{j^*,j'}\leftarrow0$, $m_{i',i^*}\leftarrow0$ and $m_{i',j^*}\leftarrow0$ for all $i'$ and $j'$.
							\end{enumerate}
				\end{enumerate}
				\item Permutation:
				\begin{enumerate}
						\item For $p$ =  1 to $P$,
						\begin{enumerate}
								\item Permute the class labels.
								\item Re-calculate $\deltav_{(i,j)}^p$ and $\wh{\Sigma}_{(i,j),(i,j)}^p$.
								\item Calculate $m_{i,j}^p = \deltav_{(i,j)}^{p\prime}\left(\wh{\Sigma}_{(i,j),(i,j)}^p\right)^{-1}\deltav_{(i,j)}^p$ for each $(i,j)\in\Pc$. 								
						\end{enumerate}
						\item Calculate the $p$-value $p_{(i,j)}$ for each $(i,j)\in\Pc$.
						\item Sort the list of pairs according to $p_{(i,j)}$ in an increasing order.
				\end{enumerate}
				\item Selection:
				Choose the top $t$ pairs which controls the false discovery rate (FDR) or according to the budget of the researcher.
\end{enumerate}

\section{Computational Issues}\label{sec:computing}
The SigJEff method considers effects beyond marginal ones. In the SigJEff framework, there are two main parts which can be computational intensive. One is to compute the statistics corresponding to all $d(d-1)/2$ pairs during the partition stage. The other is the time for permutation. To compare all the $d(d-1)/2$ pairs of variables with their permuted counterparts, the corresponding computational cost could be $(d-1)/2$ times of that of the more efficient marginal methods. For SigJEff, we significantly reduce the computational cost for the permutation stage, by permuting and calculating only the statistics for the (reduced) $\left\lfloor d/2\right\rfloor$ pairs of variables. Hence at the permutation stage, the computational cost is at the same order as the marginal method.

At the partition stage, we would have to calculate all $d(d-1)/2$ pairs in order to achieve a partition of the variables. However, when $d$ is very large, say greater than thousands, this could still cost a lot of time and memory. In this case, we propose to adopt a reasonable assumption to avoid computing all pairs.

\begin{assume}\label{assume1}
Within any nonempty subset of the whole $d$ variables that has size $d^*\leq d$, the pair with the highest Mahalanobis distance between classes, is constituted by two variables whose absolute two-sample $t$ values are ranked at the top $d_0$ among the $d^*$ variables.
\end{assume}

Assumption 1 above implies that for any collection of variables, the best pair within this collection should be searched among the best $d_0$ variables in the sense of marginal effects. Note that the assumption holds for all nonempty subset of the $d$ variables, instead of only the case where $d^*=d$. When $d_0=2$, then the best pair is also the two best variables with the highest marginal effects, which means that the ranking of the joint effect coincides with that of the marginal effect. This is opposite to what motivates this article and not what we want to assume. However, when $d_0$ is reasonably large (say hundreds), it allows the joint effect to give a different ranking of variables from the marginal effect ranking. On the other hand, when $d_0\ll d$, there is a huge save in computation since we do not need to search the best pair by calculating all pairs; instead, we can focus on a subset where the best pair is more likely to appear.

Making use of Assumption 1, we propose a faster computational strategy, where we calculate pairs incrementally. In particular, to find a best pair among the remaining variables each time, we focus on a small group of $d_0$ variables. Such strategy can be summarized as follows.

\noindent\textbf{Fast computational strategy for partition:}
\begin{enumerate}
		\item Sort variables in a descending order based on the absolute two-sample $t$ values, and save them as a waiting list $\Wc$.
		\item Active set $\Ac\leftarrow \set{\mb{pairs of the top $d_0$ variables from $\Wc$}}$; Output list $\Sc\leftarrow \emptyset$; The top $d_0$ variables are deleted from $\Wc$.
		\item Calculate statistics for all $d_0(d_0-1)/2$ pairs in $\Ac$.
		\item Promote the best pair in $\Ac$ to the end of list $\Sc$. Delete any pair from $\Ac$ whose component variable is any one of the variables in the best pair that is just promoted.
		\item Move the next two variables from $\Wc$ to $\Ac$. Calculate the new pairs created by the addition and save the pairwise results in $\Ac$.
		\item Repeat 4--5 until no variable is left in $\Wc$ and all the pairs in $\Ac$ have been promoted to $\Sc$.
\end{enumerate}

\begin{figure}[!htb]
		\includegraphics[width=\linewidth]{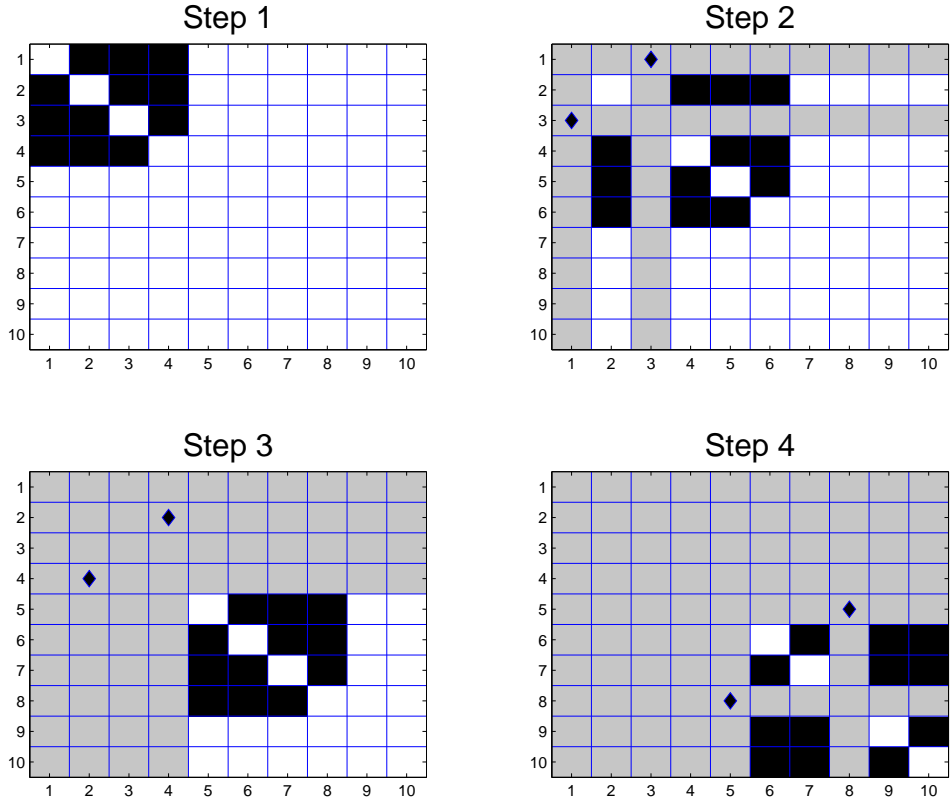}
		\caption{A toy example where $d=10$ and $d_0=4$ showing advantage of the new strategy. In each step, only 6 (4 choose 2) pairs are considered. Overall, only 21 pairs have been calculated, compared to 45 pairs without Assumption 1.}
		\label{fig:fast}
\end{figure}

Figure \ref{fig:fast} shows a toy example with $d=10$ and $d_0=4$. The subset size $d^*$ in Assumption 1 takes value 10, 8, 6 and 4 in the four steps that we show here. Before the analysis is started, all the variables are sorted according to the marginal effect. In step 1, our algorithm starts with the top 4 variables (\textit{i.e.} 6 pairs corresponding to the black cells in the figure). At step 2, pair $\set{1,3}$ (shown as diamonds) are found to be the best pair and variables $1$ and $3$ are promoted to the output list. Any pair that contains either variable 1 or variable 3 will no longer be considered (gray cells in the figure). Variables 5 and 6 substitute variables 1 and 3. Moreover, 5 new pairs ($\set{5,6}, \set{5,2}, \set{5,3}, \set{6,2}, \set{6,3}$) joins $\set{2,4}$, the pair that was not grayed out. At step 2, we still compare only  6 pairs. The procedure is repeated until each variable has been considered. In the end, we have calculated for only $21(=6+5+5+5)$ pairs. If we do not use Assumption 1 and this fast strategy, we will have to calculate for $45(=\mb{10 choose 2})$ pairs. Lastly, the first six variables in the output list of the algorithm are (1,3,2,4,5,8), which is different from the ranking by $t$ test, \textit{i.e.} (1,2,3,4,5,6). They do not appear to be substantially different in this very small toy example. However, an instance with $d=5000$ and $d_0=200$ can have two apparently very different rankings between our method and a marginal method. Despite the differences, these two rankings would have some close underlying connection, due to Assumption \ref{assume1}.

In the algorithm above, the waiting list $\Wc$ is of length $d$, which is inevitable, even for a marginal method. But the active set $\Ac$ is only of length $d_0(d_0-1)/2$ and the output list $\Sc$ is only of length $\left\lfloor d/2\right\rfloor$. This algorithm fully utilizes the advantage created by Assumption \ref{assume1}. The main computational cost is on the memory management: index, search, save and deletion. The advantage is that at any time we only need memory of size $O(d+d_0^2)$ and the number of pairs we need to calculate is $d_0(d_0-1)/2+\left\{1+2(d_0-2)\right\}\times \left\lfloor (d-d_0)/2\right\rfloor = O(d_0\times d)$, compared to $O(d^2)$ pairs to calculate without the strategy. The computational saving comes from: (1) focusing on a small group in each iteration and (2) deleting those variables which are promoted at the early stage.

In the simulation study (Section \ref{sec:simulation}) we have tried the regular SigJEff for smaller dimensions ($d=500$) and the fast SigJEff discussed in this section for higher dimensions ($d=5000$). For the real data application, we use the fast SigJEff algorithm. We have chosen $d_0=200$ for a reasonable computational time. The larger $d_0$ is, the less restriction Assumption \ref{assume1} imposes on our algorithm.

\section{Simulations}\label{sec:simulation}
We consider three different simulation settings, where the covariance structures follow $AR(1)$ process, Block-diagonal covariance and Independence (diagonal) covariance matrices. The details of the settings are given in Section \ref{sec:settings}. The methods of comparison and the measures of performance are explained in Section \ref{sec:mop}. The results are fully elaborated in Section \ref{sec:results}.

\subsection{Settings}\label{sec:settings}
\begin{example}\label{example1}[\textbf{AR(1) Process}]
This example includes $d$ variables and 100 observations (50 from each class), where $d=500$ and $5000$ respectively. For each observation (sample), we generate a $d$-long stationary AR(1) process with marginal standard deviation 1 as the $d$ variables. The first order AR coefficient (equivalently, the correlation between two adjacent variables) equals  $-0.8$. We then add mean differences to the first 50 variables, so that the squared marginal mean differences between the two classes linearly decrease to zero from variable 1 to variable 50. That is, we add $c(\sqrt{50},\sqrt{49},\cdots,\sqrt{1})^T$ to the first 50 dimensions of each observation from the first class.
 The constant $c>0$ is chosen to make the Mahalanobis distance between the two population means (a notion of the signal level) to be 2.5.
\end{example}
\begin{example}\label{example2}[\textbf{Block Diagonal Covariance}]
The dimensions and sample sizes for this example are the same as Example \ref{example1}. Define a $10\times 10$ symmetric matrix $\Sigma_0 = [\sigma_{ij}]_{i,j=1,\cdots,10}$, where $\sigma_{ii}=1$ and $\sigma_{ij}=0$, except that 4 randomly selected off-diagonal entries in the upper triangular part of $\Sigma_0$ are assigned to be $-0.8$. The lower triangular part is updated accordingly due to symmetry. Then we add $\delta = |\min(\lambda_{\mb{min}}(\Sigma_0),0)|+0.05$ to the diagonal entries of $\Sigma_0$ to make it invertible where $\lambda_{\mb{min}}(\Sigma_0)$ is the smallest eigenvalue of $\Sigma_0$, and then rescale each entry of $\Sigma_0$ by dividing them by $(1+\delta)$. Let $\Sigma$ be a $d\times d$ identity matrix, except that the first 5  diagonal blocks, each of which is $10\times 10$, all equal to $\Sigma_0$, \ie, $\Sigma = \mb{Block-Diag}\{\Sigma_0,\Sigma_0,\Sigma_0,\Sigma_0,\Sigma_0,\Id_{d-50}\}.$ Then the data vectors from each class are generated according to multivariate normal distributions $\xv_{\pm,i}\stackrel{iid}{\sim} MVN_d(\muv_\pm,\Sigma)$, where $\muv_+=c(\sqrt{50},\sqrt{49},\cdots,\sqrt{1},0,0,\cdots,0)^T$, $\muv_-=\0v$ and the constant $c>0$ is chosen to make the Mahalanobis distance between the two population means equal 2.5.
\end{example}
\begin{example}\label{example3}[\textbf{Independent Covariance}]
This example is the same as Example \ref{example2} except that the covariance matrix is $\Id_{d}$. Note that in this setting, there is \textit{only} marginal effect and no joint effect is included.
\end{example}

\subsection{Methods and measures of performance}\label{sec:mop}
First, we compare the number of true non-null variables that are selected among the selected variables from SigJEff, SAM, and LASSO over 100 runs of simulation. The more true non-null variables, the better the variable selection result is as the fewer true variables is missed.
Second, we compare the average false discover proportion (FDP) of the variable selection results from the three variable selection procedures, over different numbers of variables selected. FDP is defined as the proportion of the true null variables among all the selected variables. We would like to have small  FDPs.

Besides the previous measures, we also compare the misclassification rates using the selected variable sets. We apply  two standard classifiers, Support Vector Machine \citep[SVM;][]{Cortes1995Support,Vapnik1999nature} and Linear Discriminant Analysis \citep[LDA;][]{Fisher1936use} to test data sets based on the selected variable selection sets using SigJEff and SAM. The test data sets are generated in the same way as the training data sets and of 10 times larger. For LASSO, as it can be used as a binary classification method, its misclassification error rates is reported directly. Cross validations are used for the parameter tuning in SVM. We use the SVM implementation by \texttt{R/e1071} and the default setting therein for tuning. We use \texttt{glmnet} to train LASSO after relabeling the response variables $y\leftarrow n/n_1$ for $y=+1$ and $y\leftarrow -n/n_2$ for $y=-1$, where $n_1$ and $n_2$ are the sample sizes of the two classes. We use the empirical version of SigJEff $p$-value in simulations, as in our simulation settings the effect is not strong and therefore using the empirical $p$-value is sufficient.

\subsection{Simulation results}\label{sec:results}
\subsubsection{AR(1) Process}
\begin{figure*}[ht]
		\includegraphics[width=0.8\linewidth]{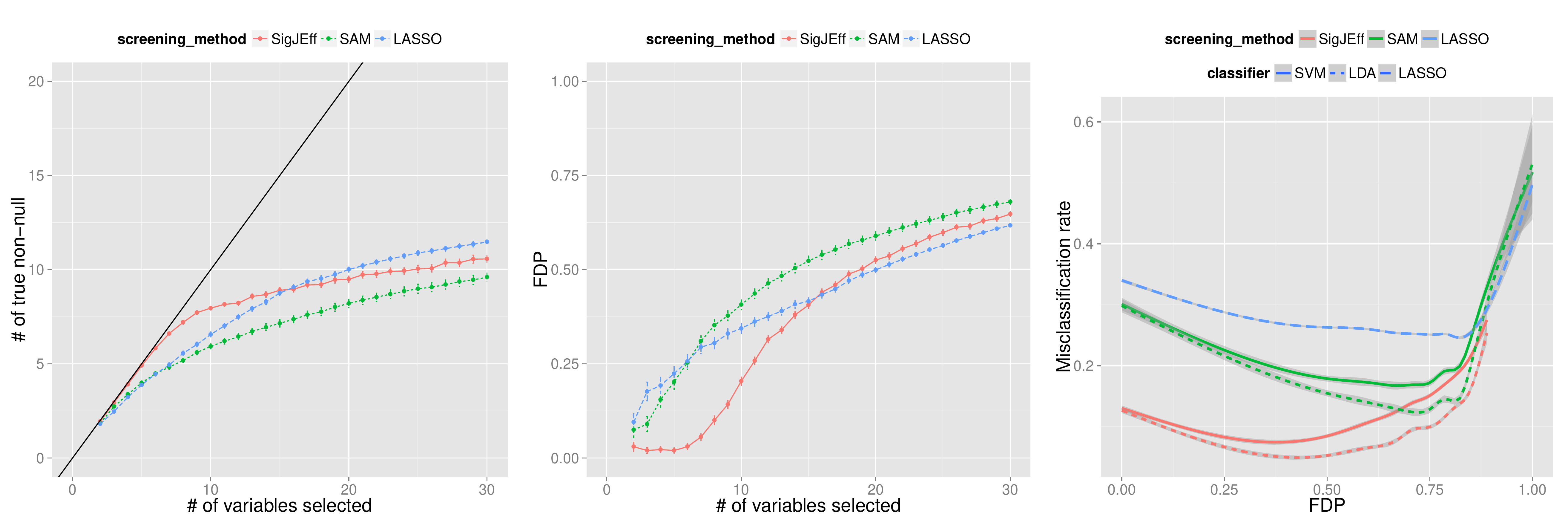}
		\includegraphics[width=0.8\linewidth]{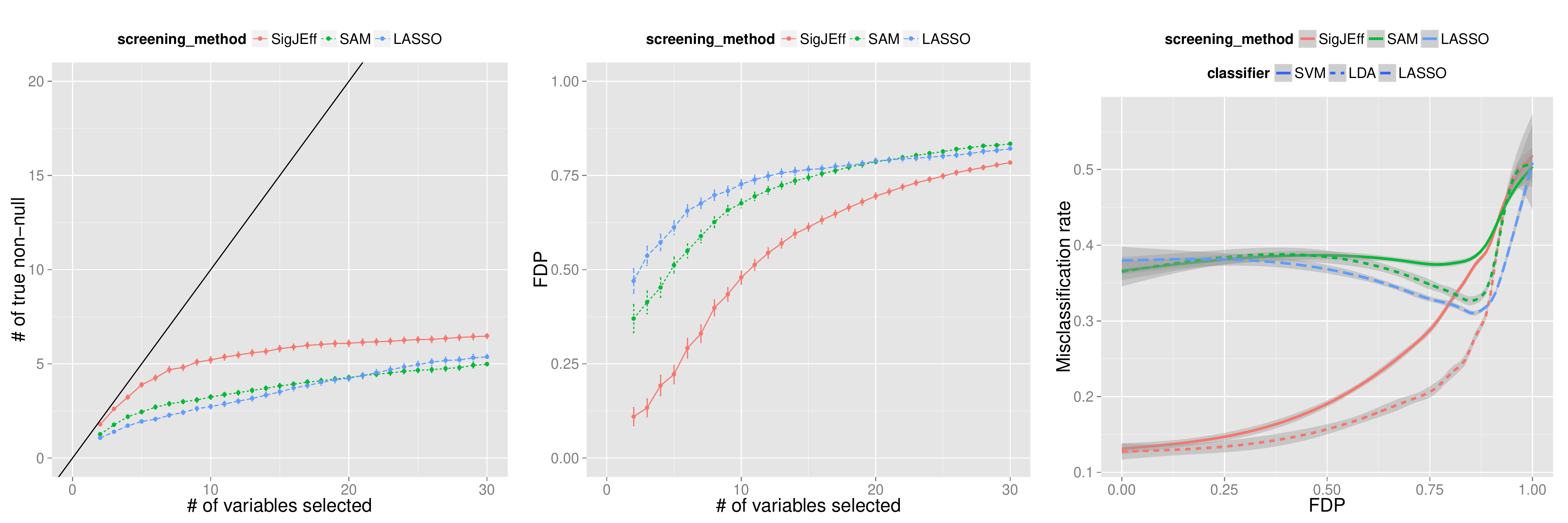}
		\caption{AR(1) Process example: The number of true non-null as function of the number of variables selected, the false discover proportion (FDP) as function of the number of variables selected and the misclassification rate for test data as function of FDP, from SigJEff, SAM and LASSO. The top row is for the low dimensional setting and the bottom row is for the high dimensional setting. The results show that SigJEff selects more true non-null variables than SAM and LASSO, gives better variable selection quality (lower FDP), and when given the same FDP, the submodels chosen by SigJEff can give better classification performance. It also shows that the SigJEff method enjoys greater advantage when the dimension increases.}
		\label{fig:expm1}
\end{figure*}

In Figure \ref{fig:expm1}, we plot the numbers of true non-null among the variables that are selected as functions of the total number of variables selected (left panel), the FDPs as functions of the number of variables selected (middle panel) and the misclassification rates for test data set as functions of FDP, from SigJEff, SAM and LASSO (right panel). The top row is for the low dimensional example ($d=500$) and the bottom row is for the high dimensional example ($d=5000$). The results in the low dimensional example show that SigJEff selects more true non-null variables than SAM uniformly and than LASSO when the number of variables selected is fewer than 16. Similarly, SigJEff gives better variable selection quality (lower FDP) than SAM over the broad, and than LASSO for smaller submodel sizes. However, note that LASSO gives better FDP only when the submodel size is large, in which case the FDP value is relatively high (about 40\%). For the high dimensional example, the advantage of SigJEff is even larger and it is better than SAM and LASSO uniformly for the experiments we studied.

\begin{figure*}[ht]
		\includegraphics[width=0.8\linewidth]{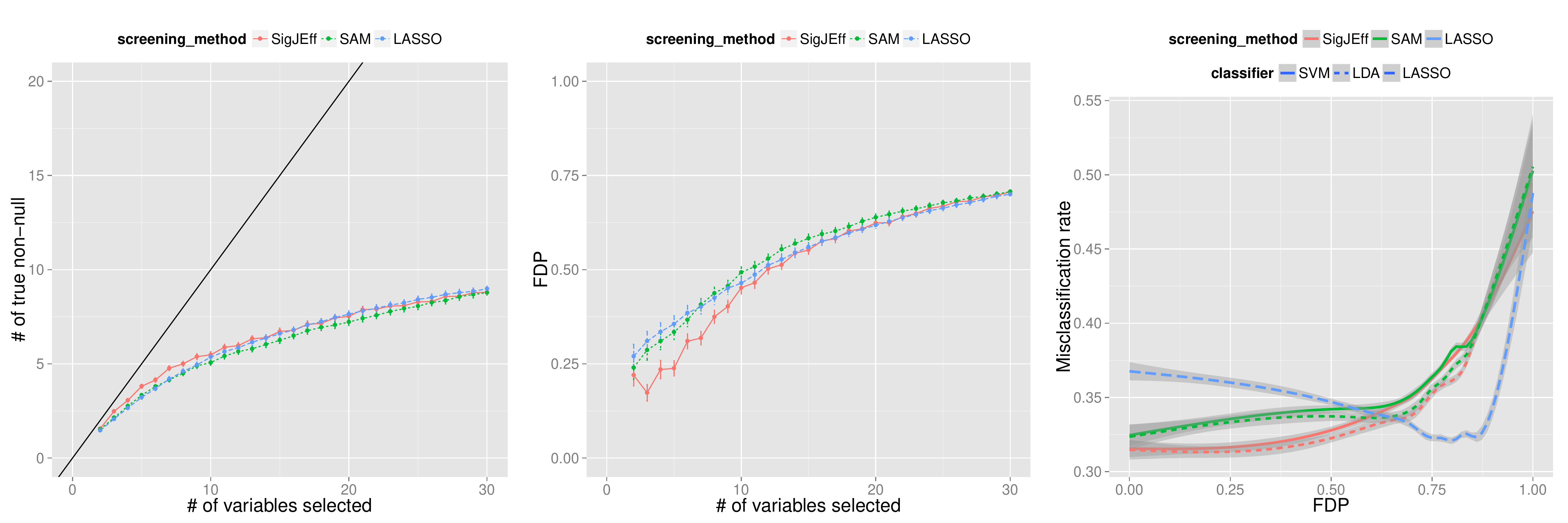}
		\includegraphics[width=0.8\linewidth]{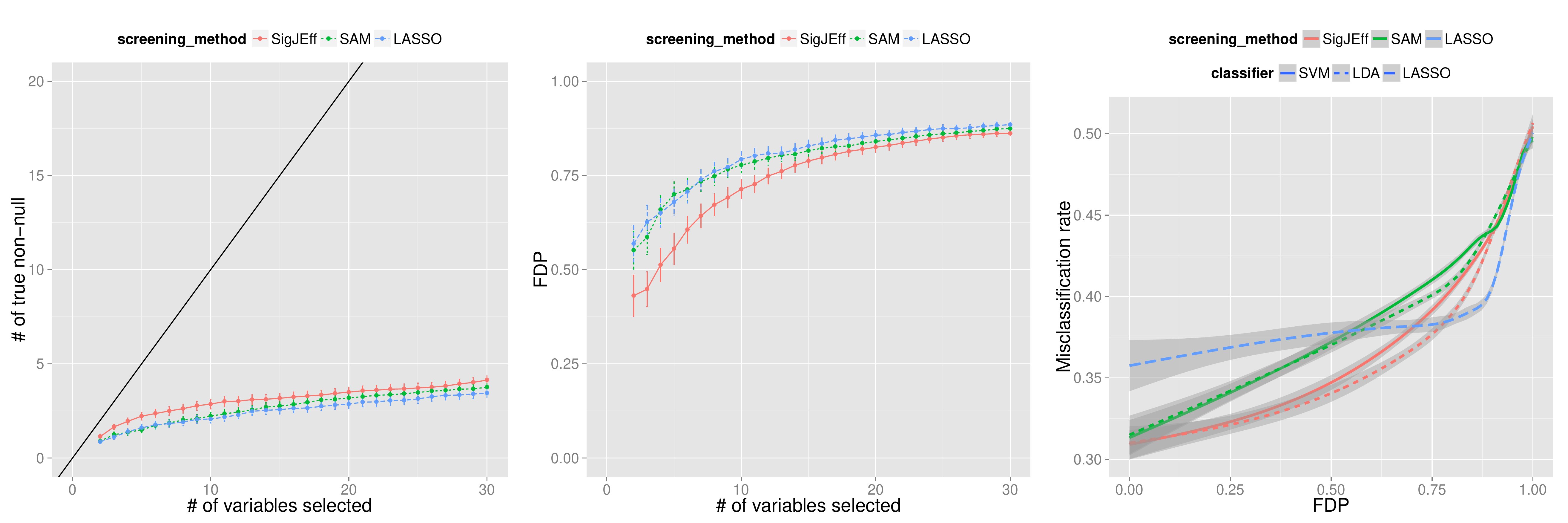}
		\caption{Block Diagonal example: The number of true non-null as function of the number of variables selected, the false discover proportion (FDP) as function of the number of variables selected and the misclassification rate for test data as function of FDP, from SigJEff, SAM and LASSO. The top row is for the low dimensional setting and the bottom row is for the high dimensional setting.  The results show that SigJEff selects more true non-null variables than SAM and LASSO, gives better variable selection quality (lower FDP), and when given the same FDP. The submodels chosen by SigJEff can give better classification performance when FDP is relatively small. The margins between SigJEff and the other two methods are not as large as those seen in the previous example. As the dimensions increase, the advantage of SigJEff over its competitors does not seem to be increased.}
		\label{fig:expm2}
\end{figure*}
\begin{figure*}[ht]
	\centering
		\includegraphics[width=0.8\linewidth]{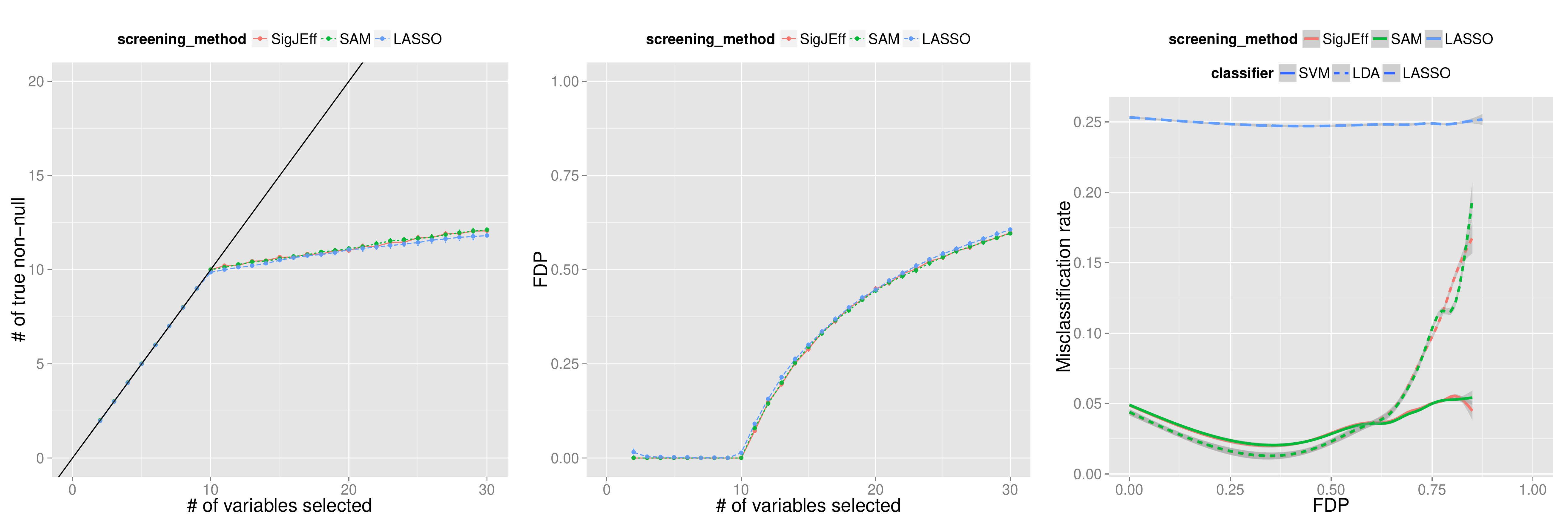}
		\includegraphics[width=0.8\linewidth]{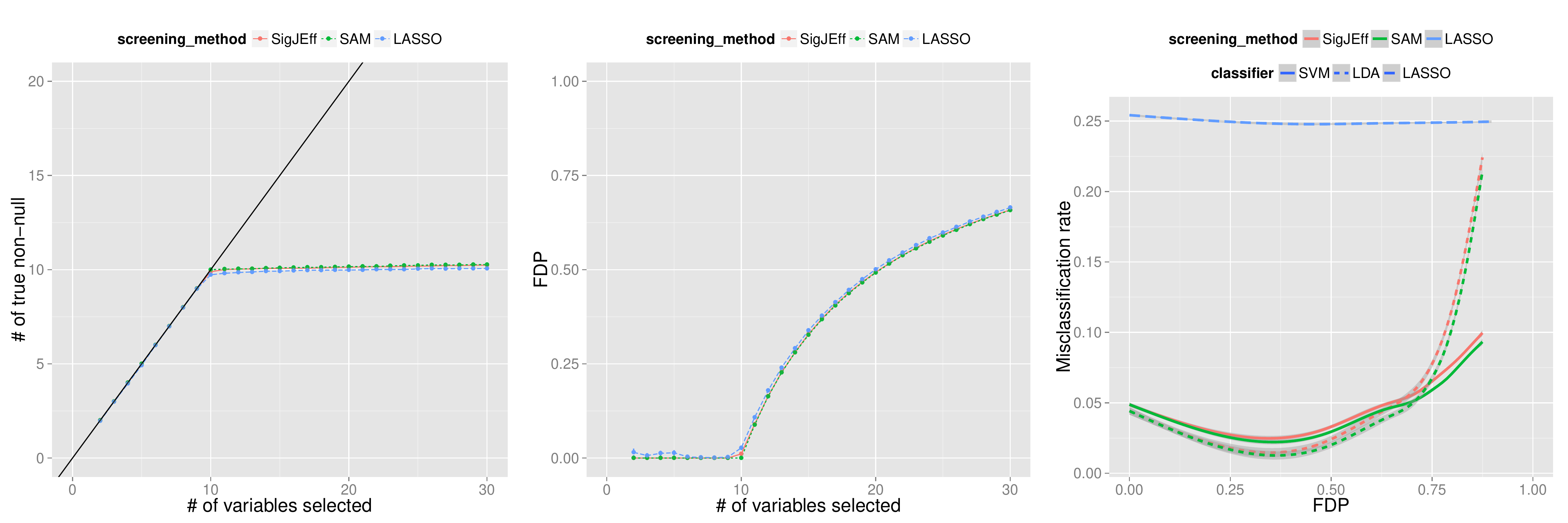}
	\caption{Independent example: The number of true non-null as function of the number of variables selected, the false discover proportion (FDP) as function of the number of variables selected and the misclassification rate for test data as function of FDP, from SigJEff, SAM and LASSO. The top row is for the low dimensional setting and the bottom row is for the high dimensional setting.  The results show that the variable selection and classification performance for SigJEff and SAM are almost identical. The variable selection performance of LASSO is just a little worse than them.}
	\label{fig:expm3}
\end{figure*}
\begin{figure*}[ht]
		\includegraphics[width=0.8\linewidth]{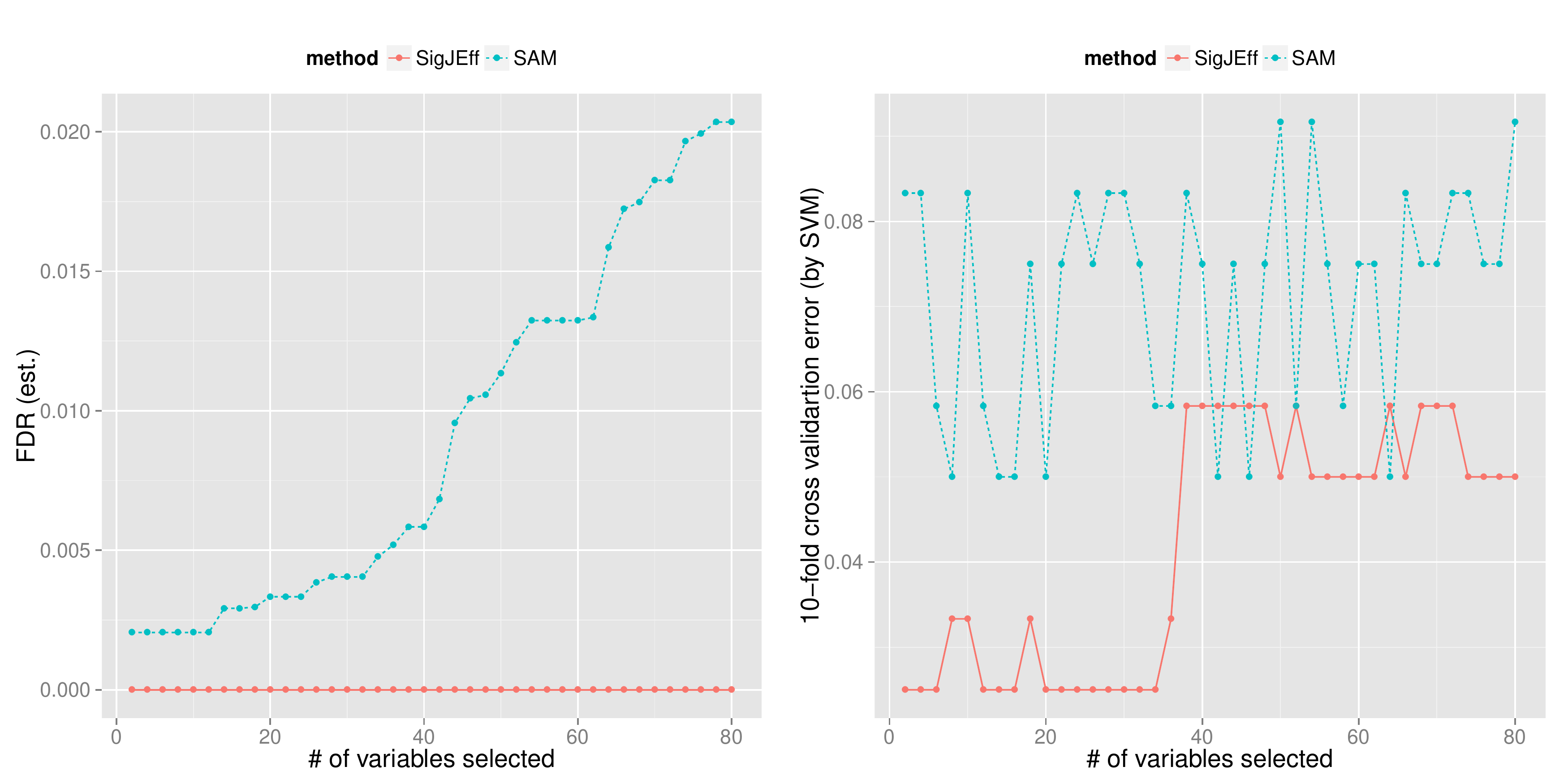}
		\caption{The estimated false discovery rate and the 10-fold cross validation error (by SVM) based on variable sets selected by SigJEff and SAM respectively. The estimated SigJEff FDR is always zero. Note this can be due to the different ways of FDR estimation. More informative (and reliable) is the classification performance shown in the right panel. SigJEff dominates SAM in terms of classification when the SVM classifier is used.}
		\label{fig:real1}
\end{figure*}

Lastly, given the same FDP, the submodels chosen by SigJEff can give the best classification performance.
In particular, as we can see in the right panel, the misclassification rates from the SVM classifier are depicted using solid curves and those from the LDA classifier are depicted using the dashed curves. The LASSO misclassification rates are shown using the blue longdashed curve. We can see that the SigJEff-SVM classifier (\ie, SigJEff variable selection followed by SVM classifier) is better than the SAM-SVM classifier  uniformly over different FDPs; the  SigJEff-LDA classifier is better than the SAM-LDA classifier uniformly over different FDPs. All four classifiers above are better than LASSO in terms of misclassification rate. Again, the advantage of SigJEff in the high dimensional example is more obvious than that in the low dimensional example.

\subsubsection{Block Diagonal covariance}
Similar results can be seen in Figure \ref{fig:expm2}, though the margins between SigJEff and the other two methods become smaller than those in the last example (at least in the low dimensional example).

The numbers of true non-null chosen by all three methods are fairly close, with SigJEff leading slightly when the submodel sizes are small (the advantage is visually separable as in the high dimensional example in the bottom row). The LASSO submodels are similar to SAM for smaller submodel sizes and are similar to the SigJEff submodels when the submodel sizes increase. The gain of FDP seems to be more substantial. In the middle panel,
SAM and LASSO are very close in FDP, while SigJEff submodels enjoy smaller FDPs. Lastly, the submodels chosen by SigJEff can give better classification performance when FDP is relatively small. When FDP is as large as 75\% to 90\%, the classification performances from different variable selection methods and classifers are mixing. Note that when the variable selection quality is so low, one may not care about the classification performance as much. For small FDP, the SigJEff method is clearly better than SAM, which is better than LASSO. Note that the best classification performance of LASSO is attained when the FDP is about 80\%. Overall, the best classification performance is attained by the two SigJEff related classifiers.

\subsubsection{Independent covariance}
The Independent covariance example is meant to be an example where SigJEff and SAM should perform similarly because there is no true additional joint effects besides marginal effects due to the independence setting. The simulation results validate the conjecture. In Figure \ref{fig:expm3}, we can see that the variable selection quality of SAM and SigJEff are almost identical, while the LASSO variable selection is worse by a very narrow margin.

In terms of classification, again, SigJEff and SAM perform almost identically. The performance of LASSO is not as good, which is probably due to the fact that LASSO is not tailored for classification.

The results from this example suggest that, in practice, even if there was no joint effect, using SigJEff would not give results that are worse than using marginal methods.

\subsection{Remarks}
In the left and middle panels of each figure, we use error bars to depict the standard error of the estimated mean number of true non-null and mean FDP.

In the right panel, for better visualization and  presentation, nonparametric smoothing is applied to fit the conditional mean of the `misclassification rate' as a function of `FDP'. See \texttt{geom\_smooth} function in the \texttt{R/ggplot2} package for details. The standard error is shown as the shadow using the default setting of the smoothing function.

The purpose of this set of simulation study is to understand the performance of SigJEff, SAM and LASSO over a broad range of submodel sizes (\ie, the number of variables selected varies). In the simulations, we do not apply a threshold to choose the best submodel size, because
\begin{enumerate}
		\item The choice of the submodel size depends on the budget and the capacity of the researcher;
		\item Studying the performance for a particular choice of the submodel size would only give us a comparison from one single aspect, while what we have done here is to try to look at the whole picture and to understand the methods under investigation for different submodel sizes.
\end{enumerate}

We would like to point out that although one can apply SVM or LDA to the variable set selected by LASSO and evaluate the classification performance, our focus is to first select variables and then apply classification procedures, such as SVM or LDA.

Lastly, the simulation study was run in parallel in a cluster of about 300 computers whose average speed is 18800 MIPS and average memory is 1.78 GB. In the high dimensional examples, the average CPU time for each run of the SAM procedure is 0.215, 0.214 and 0.165 seconds respectively for the three settings, compared to 251.23, 205.59 and 157.46  seconds for the SigJEff procedure. Although the computational time for SigJEff is much longer than that for SAM, it is quite efficient overall. Furthermore, we believe that there is still room for improvement such as through parallelization.

\section{Real data application}\label{sec:realdata}
Recurrent genomic abnormalities were cataloged by the Cancer Genome Atlas Research Network in the glioblastoma multiforme (GBM) data sets. \citet{Verhaak2010Integrated} classified GBM into four subtypes: \textit{Proneural}, \textit{Neural}, \textit{Classical} and \textit{Mesenchymal}. We focus on the \textit{Proneural} subtype in this article. It is found that point mutations in the gene IDH1 appeared in the \textit{Proneural} data set. For the purpose of classification, we define two classes based on the status of IDH1, being with and without mutations. The sample size is 37 (with 11 mutations, and 26 no-mutation). This data set has 11338 genes.

The sample size of the data is really small, compared to the dimensions. Moreover, it turns out that the signal in this data set is so strong that hundreds of genes are significant. In order to conduct our study and compare the performance of SigJEff and SAM as means of significance analysis, we first screen the variables from the bottom by deleting all those variables whose marginal overall standard deviations are less than or equal to 0.5. Here the threshold 0.5 is an ad hoc choice. However, it helps to throw away non-informative variables. Note that this is an unsupervised screening which does not employ any class information. After the pre-screening step, the data set remains with 4280 dimensions.

We apply SigJEff and SAM to the dimension-reduced data set. In SigJEff, we use the robust Gaussian fit version of the $p$-value because we expect to have some genes with very strong effect. As shown in Figure \ref{fig:real1}, we first estimate the false discover rate (FDR) of SigJEff and SAM for different submodel sizes and compare the cross validation error of SVM based on the submodels given by the SigJEff and the SAM procedures. The estimated SigJEff FDR is always zero. Note this can be due to the different ways of FDR estimation. Here, we are not claiming that the SigJEff can control FDR better than SAM, although this seems to be a reasonable conjecture from the simulations in the previous section. More informative (and reliable) conclusion can be drawn from the classification performance shown in the right panel of Figure \ref{fig:real1}. SigJEff dominates SAM in terms of SVM classification over different submodel sizes.

\begin{figure*}[!ht]
		\includegraphics[width=0.8\linewidth]{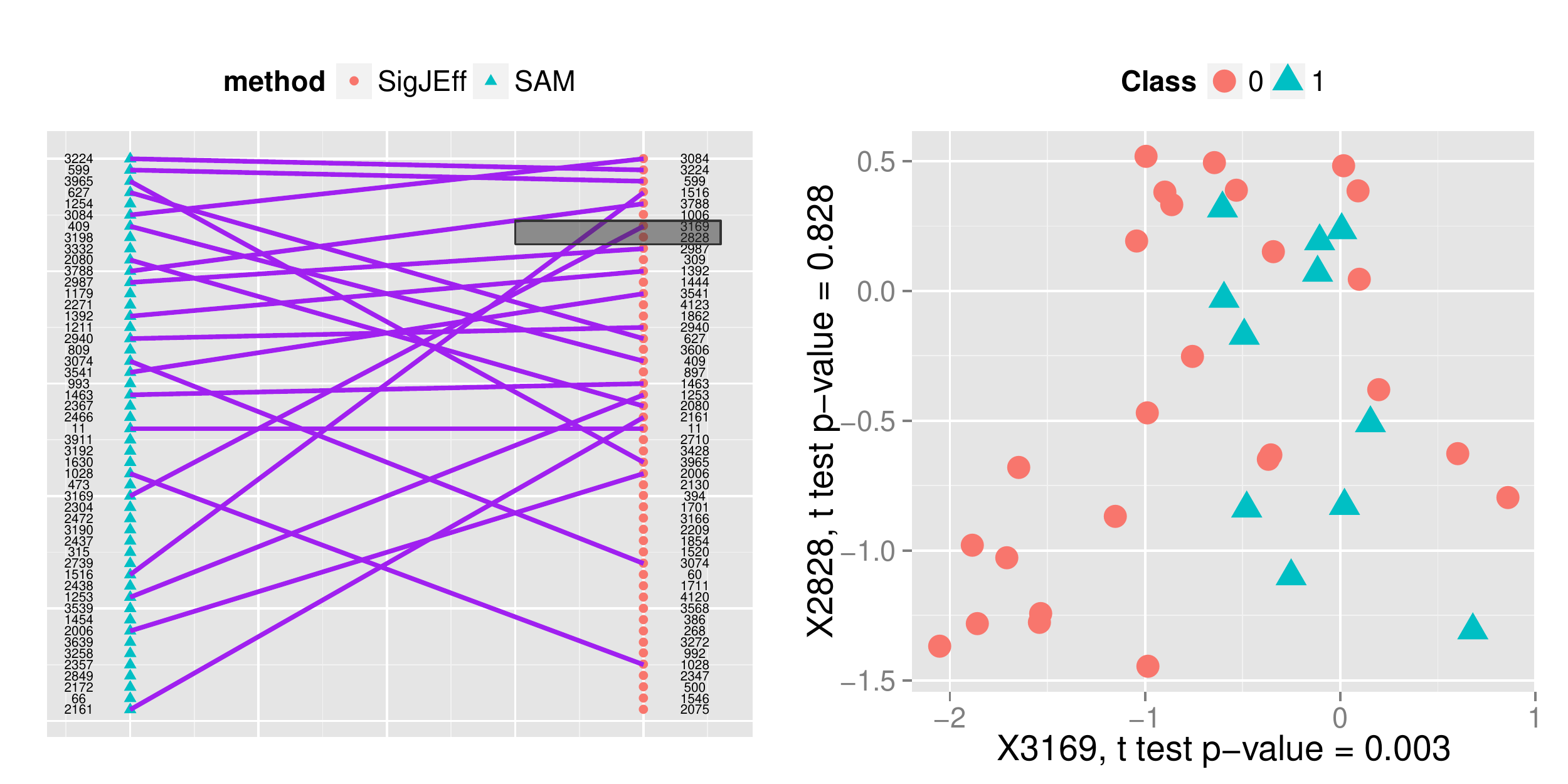}
		\caption{Left panel: The gene indices of the first 50 genes for the real data set in Section \ref{sec:realdata} that are selected by SigJEff (on the right) and by SAM (on the left) respectively. Shows the different ranks of variables by the two methods. Right panel: the scatter plot of the samples from the two classes based on gene \#3169 and gene \#2828. The latter is not SAM important at all (large $t$ test $p$-value), while the former is among the top 50 most SAM-important, but barely makes the top 30. When viewing these two variables together, a pattern can be seen that the Class `1' samples are around the southeast corner of the plot while the Class `0' samples are at the northwest. Such a pattern is not visible for each individual variable.}
		\label{fig:real2}
\end{figure*}

The good performance of SigJEff in this real data application motivates us to carefully check the genes that are selected by each method. In the left panel of Figure \ref{fig:real2}, we list, from top to bottom, the gene indices of the first 50 genes that are selected by SigJEff (on the right) and by SAM (on the left) respectively. Common genes selected by both methods are connected by a purple line segment. This plot shows the different rankings of variables selected by SigJEff and SAM. The most important SAM genes, \#3224 and \#599, are viewed important by SigJEff as well. However, there are some SigJEff-important genes that are not recognized by SAM, for example, \#1516, \#1006, and so on. We now take a deeper look at the gene pair, \#3169 and \#2828. The latter is not SAM-important at all, while the former is among the 50 most SAM-important genes, but barely makes the top 30 list. When they are paired together, they are more important and obtain a better rank  by SigJEff. In the right panel of Figure \ref{fig:real2}, we show the scatter plot of the data based on these two variables. The two-sample $t$-test $p$-value of gene \#3169 is 0.003, which explains why it makes the top 50 list of SAM-important genes. Note that gene \#2828 only has 0.828 $t$-test $p$-value, which is the reason that SAM does not select it. However, when these two variables join together, it appears from the scatter plot that the Class `1' samples are around the southeast corner of the plot while the Class `0' samples are at the northwest. It is the joint effect like this that drives the improvement of the classification performance when we use SigJEff to select variables.

\noindent\textsc{Remark:} In Figure \ref{fig:real1}, we have estimated the FDR of the variable selection set of SigJEff. Our FDR estimation follows the standard procedure and is parallel to the method employed by SAM. The details can be found in the appendix. 
\section{Discussion}\label{sec:Conclusion}
In this article, we propose a  simple and useful procedure to perform pairwise variable selection via assessing
joint effects useful for classification. We use a permutation procedure to select pairs of variables. Although this procedure is not as fast as the marginal methods such as SAM or Sure Independent Screening \citep{Fan2008Sure}, it is relatively efficient and more importantly, it can help to understand different aspects of the data which the marginal methods do not cover. Our numerical study shows that one may not lose much  by using pairwise variable selection even when there is no true joint effect, because variables that are marginally significant are usually pairwise significant as well. 

For ultra-dimensional data, there are a few strategies that may be taken to speed up the computing. First, one may keep those variables that are extremely strong first and leave them out of the SigJEff procedure, since these variables will likely be picked up by SigJEff anyway. SigJEff would work better for those variables in the gray zone where variables are moderately strong, but not strong enough to be called by a marginal method. One can also delete those variables that seem to be non-informative, such as those with very small variation. In this article, we have also considered an assumption which regulates the SigJEff partition step, and can help to speed up significantly on computation and memory.

Our method may be extended to more than two variables, by assessing the Mahalanobis distances between the two classes on more than two variables. The corresponding computation cost will be higher  and furthermore, the procedure and the corresponding interpretation becomes much more involved. Thus it does sound appealing, we have not extend our research toward that end yet.

Although our focus is on classification, SigJEff can be extended to accommodate the regression setting. In that case, one needs to define a proper criterion for the variable partition and a statistic which measures the correlation between the variable set and the response variable. Further investigation is needed.

It is worth noting that selecting the pairwise joint effect for classification is not the same as selecting the interaction effect. For the latter, see \citet{bien2012lasso} and the references therein. In particular, in the context of classification, selecting interaction focuses on the nonlinearity of the discrimination function and the classification boundary. On the other hand, our SigJEff test implicitly assumes linear boundary due to the use of the Mahalanobis distance as the statistic, although one could apply nonlinear classification methods to the resulting variables selected by SigJEff.

Note that since we rank $d/2$ disjoint pairs, correlation is not a big issue for the FDR procedure. However, even with the simplification, correlation between pairs may still exist. Thus, a better multiple comparison adjustment procedure such as \cite{Fan2012Estimating} can be helpful.

Finally, we would like to point out that there is a large literature in recent years on the use of sparsity for variable selection, such as LASSO, SCAD \citep{Fan2001Variable}, DS \citep{Candes2007Dantzig} \textit{etc}. Our proposed SigJEff is not intended to replace such sparse penalized methods, instead, we suggest to use SigJEff for prescreening and then apply one of these variable selection methods post the SigJEff procedure, much in the same spirit as SIS-SCAD, SIS-DS, \textit{etc.} as proposed in \citet{Fan2008Sure}. We expect a combination of SigJEff screening with sparse penalized methods after screening can lead to accurate prediction and selection.

The software of SigJEff can be found on the corresponding author's website: \url{http://www.math.binghamton.edu/qiao}.

\section*{Appendix}
\noindent\textbf{Estimation procedure of FDR for SigJEff.}
\begin{enumerate}
	\item For each cutoff value $c>0$, compute the total number of significant pairs from the original data, \ie, $\#\set{m_{i,j}>c:~(i,j)\in\Pc}$ , and the median number of pairs called significant, by computing the median number of $m_{i,j}^p$ values among each of the $P$ sets of $\left\lfloor d/2\right\rfloor$ pairs of variables, that fall above $c$. Similarly for the 90th percentile of pairs called significant.
	\item Estimate $\pi_0$, the proportion of true null pairs in the data set as follows:
	\begin{enumerate}
		\item Compute $q50$, the median of all the permuted statistics, $m_{i,j}^p$. Note that there are $P\times\left\lfloor d/2\right\rfloor$ such values.
		\item Compute $\wh\pi_0=\#\set{m_{i,j}<q50}/(0.5\left\lfloor d/2\right\rfloor)$, where $m_{i,j}$ is the statistic from the original data set and there are $\left\lfloor d/2\right\rfloor$ such values.
		\item Truncate $\wh\pi_0$ at 1: $\wh\pi_0\leftarrow \min(\wh\pi_0,1)$.
	\end{enumerate}
	\item The median and 90th percentile of the number of pairs called significant from Step 1 are multiplied by $\wh\pi_0$ to obtain estimations of the median and 90th percentile of the number of \textit{falsely} called pairs.
	\item The SigJEff FDR is computed as [the median (or 90th percentile) of the number of falsely called pairs] divided by [the number of pairs called significant in the original data].
\end{enumerate}

\section*{Acknowledgements}
We thank Katherine Hoadley, Neil Hayes and their colleagues for allowing us to analyze the
GBM data and for giving advices on the results. The research was supported in part by Binghamton University Dean's New Faculty Startup Funds, NSF grant DMS-0747575, NIH grants R01-CA149569, R01-05-010134
and P50-CA58223-15, and a grant from the Simons Foundation (\#246649).

\bibliographystyle{asa}
\bibliography{bib-sj2}

\end{document}